\pdfoutput=1
\documentclass[aps,10pt,pre,showpacs,showkeys,twocolumn,footinbib,bibnotes,longbibliography]{revtex4-1}
\usepackage{microtype}
\usepackage{graphicx}
\usepackage{amsmath}
\usepackage{amssymb}
\usepackage{color}
\definecolor{myblue}{rgb}{0.153,0.322,0.706}
\usepackage[colorlinks,linkcolor=myblue,urlcolor=myblue,citecolor=myblue,breaklinks=true]{hyperref}

\renewcommand{\emph}{\textit}

\newcommand{\be}{\begin{equation}}
\newcommand{\ee}{\end{equation}}
\newcommand{\ra}{\rightarrow}
\newcommand{\cL}{\mathcal{L}}
\newcommand{\cH}{\mathcal{H}}
\newcommand{\reals}{\mathbb{R}}

\newcommand{\p}{\partial}
\newcommand{\cD}{\mathcal{D}}
\newcommand{\hX}{\hat X}
\newcommand{\oup}{\textrm{OUP}}
\newcommand{\tF}{\tilde F}
\newcommand{\trho}{\tilde\rho}
\newcommand{\tI}{\tilde I}
\DeclareMathOperator{\erf}{erf}	
\DeclareMathOperator{\Ai}{Ai}	

\graphicspath{{../figures/cmr/}}

\begin{document}
\vspace*{0.02in}
\title{Dynamical large deviations of reflected diffusions}

\author{Johan du Buisson}
\email{johan.dubuisson@gmail.com}
\affiliation{\mbox{Institute of Theoretical Physics, Department of Physics, Stellenbosch University, Stellenbosch 7600, South Africa}}

\author{Hugo Touchette}
\email{htouchet@alum.mit.edu, htouchette@sun.ac.za}
\affiliation{Department of Mathematical Sciences, Stellenbosch University, Stellenbosch 7600, South Africa}

\date{\today}

\begin{abstract}
We study the large deviations of time-integrated observables of Markov diffusions that have perfectly reflecting boundaries. We discuss how the standard spectral approach to dynamical large deviations must be modified to account for such boundaries by imposing zero-current conditions, leading to Neumann or Robin boundary conditions, and how these conditions affect the driven process, which describes how large deviations arise in the long-time limit. The results are illustrated with the drifted Brownian motion and the Ornstein--Uhlenbeck process reflected at the origin. Other types of boundaries and applications are discussed.
\end{abstract}


\maketitle

\section{Introduction}

The use of stochastic differential equations (SDEs) for modelling noisy diffusive systems often requires that we specify the location of boundaries or ``walls'' when the system of interest evolves in a confined space or has a state that is inherently bounded \cite{gardiner1985,schuss2013,bressloff2014}. Examples include molecules diffusing in cells~\cite{bressloff2014,grebenkov2007b,holcman2017}, particle transport in porous media \cite{grebenkov2007b}, as well as diffusive limits of population \cite{bressloff2014} and queueing dynamics \cite{giorno1986,ward2003,linetsky2005} which have a positive state. In each case, one must also define what happens when a boundary is reached by specifying a boundary type or condition on the density $\rho$ and current $J$ entering in the Fokker--Planck equation \cite{gardiner1985}. Reflecting boundaries, for instance, are defined by requiring $J=0$ at the boundary, whereas absorbing boundaries, related to extinctions in population models, are such that $\rho=0$ at the boundary. Other types of boundaries are possible, including partially reflective \cite{singer2008}, reactive \cite{erban2007,chapman2016,pal2019}, and sticky \cite{harrison1981,bou2019,engelbert2014}, and arise in biological and chemical applications.

Many studies, starting with Feller \cite{feller1952,feller1954,peskir2015}, have looked at the effect of boundaries on SDEs at the level of probability distributions (time-dependent or stationary) and mean first-passage times \cite{schuss2013,bressloff2014,grebenkov2007b}. In this paper, we investigate this effect on the long-time large deviations of time averages of the form
\be
S_T = \frac{1}{T}\int_0^T f(X_t)dt,
\label{eqobs1}
\ee
where $f$ is some function of the state $X_t$ of a bounded SDE. The random variable $S_T$ can be related, depending on the system considered, to various physical quantities that are integrated over time, and is called for this reason a \emph{dynamical observable} \cite{sekimoto2010,seifert2012,touchette2017}. For simplicity, we study one-dimensional systems, so that $X_t$ evolves in a closed interval $[a,b]$ of $\reals$ and consider perfect reflections at the endpoints $a$ and $b$ \footnote{We do not consider observables involving integrals of the increments of $X_t$, as they are trivial for one-dimensional diffusions.}. Other types of boundaries are discussed in the conclusion. 

Large deviations have been studied before for reflected SDEs, in particular, by Grebenkov \cite{grebenkov2007}, Forde \textit{et al.}~\cite{forde2015}, and Fatalov \cite{fatalov2017}, who obtain the rate function of various functionals of reflected Brownian motion, including its area and the residence time at a reflecting point. Pinsky \cite{pinsky1985,pinsky1985b,pinsky1985c,pinsky1985d} and Budhiraja and Dupuis \cite{budhiraja2003} also study the large deviations of bounded diffusions, but do so at the level of empirical densities, the so-called ``level~2'' of large deviations, rather than time averages, which corresponds to ``level~1'' \cite{touchette2009}. Finally, many studies \cite{dupuis1987b,sheu1998,majewski1998,ignatyuk1994,ignatiouk2005,bo2009} consider escape-type events occurring in the low-noise limit, which fall within the Freidlin--Wentzell theory of large deviations \cite{freidlin1984}. In this case, the rare events of interest typically involve the state $X_t$ at a fixed or random time rather than time averages of $X_t$, as in \eqref{eqobs1}. 

In this work, we focus on the long-time limit and extend the studies above by deriving the reflective boundary conditions of the spectral problem that underlies the calculation of dynamical large deviations \cite{touchette2017}. Our results clarify the source of the boundary conditions used in \cite{grebenkov2007,fatalov2017} and extend them to more general SDEs and observables. We also investigate how the presence of reflecting boundaries affects the \emph{driven process}, introduced in \cite{chetrite2013,chetrite2014,chetrite2015} to explain, via a modified SDE, how fluctuations of $S_T$ away from its typical value are created in time. The main result that we obtain for this process, which is also called the auxiliary or effective process \cite{evans2004,jack2010b,jack2015}, is that its drift generally differs from the drift of the original SDE everywhere except at the boundaries, due to the $J=0$ condition which is also satisfied by the driven process.

These results can be applied to study the large deviations of many equilibrium and nonequilibrium diffusions, including manipulated Brownian particles, which necessarily evolve in a confined environment and so can interact with walls. As illustrations, we consider two simple reflected diffusions, namely, the reflected Ornstein-Uhlenbeck process, which models the dynamics of an underdamped Brownian particle pulled linearly towards a reflecting wall as well as the dynamics of queueing systems in the heavy-traffic regime \cite{ward2003}, and the reflected Brownian motion with negative drift, which models a Brownian particle pulled to a wall by a constant force such as gravity. Other applications, boundary types, and open problems are discussed in the conclusion.

\section{Reflected diffusions}

We consider a one-dimensional Markov diffusion $(X_t)_{t\geq 0}$ defined by the SDE
\be
dX_t = F(X_t)dt+\sigma dW_t,
\label{eqsde1}
\ee
which we restrict to the interval $[a,b]$ with $a<b$ and either $a$ or $b$ (or both) finite. The function $F(x)$ is called the \emph{force} or \emph{drift}, and is assumed to be such that the boundaries of $[a,b]$ are reachable in finite time from the interior of this interval (regular boundaries) \cite{karlin1981}. The constant $\sigma>0$ is the noise amplitude multiplying the increments of the Brownian motion $W_t\in\reals$, representing in SDE form a Gaussian white noise. The more general case where $\sigma$ depends on $X_t$ can be covered using the methods explained in~\cite{chetrite2014}.

To complete the model, we must specify the behavior of the process at the boundaries of $[a,b]$. Mathematically, this can be done at the level of the SDE or at the level of the Fokker--Planck equation
\be
\p_t \rho(x,t) = -\p_x F(x)\rho(x,t)+\frac{\sigma^2}{2}\p_{xx}\rho(x,t),
\ee
which governs the evolution of the time-dependent probability density $\rho(x,t)$ of $X_t$, starting from some initial density $\rho(x,0)$ for $X_0$. Rewriting this equation as a conservation equation
\be
\p_t \rho(x,t) = -\p_x J_{F,\rho}(x,t),
\ee
we identify the probability current
\be
J_{F,\rho}(x,t) = F(x) \rho(x,t) - \frac{\sigma^2}{2}\p_x \rho(x,t).
\label{current}
\ee
Perfect reflections are then imposed by requiring that this current vanish at $a$ and $b$ (approaching these points from the interior). Thus,
\be
J_{F,\rho}(a^+,t) =0= J_{F,\rho}(b^-,t)
\label{eqcurr1}
\ee
at all times $t$, where $a^+=a+0$ and $b^-=b-0$. This follows, as is well known \cite{gardiner1985}, because a reflecting boundary cannot be crossed, so the normal component of the probability current at the boundary, which in one dimension is simply the current itself, must be equal to zero. 

If $X_t$ is ergodic, we also have in the long-time limit
\be
J_{F,\rho^*}(a^+) = 0=J_{F,\rho^*}(b^-),
\label{eqstatcurr1}
\ee
where  $\rho^*(x)$ is the unique stationary distribution of the Fokker--Planck equation \footnote{There is no time variable at this point because we are dealing with the stationary regime.}. For one-dimensional diffusions, we have in fact $J_{F,\rho^*}(x)=0$ not just at the boundary points but for all $x\in [a,b]$, since the stationary current is constant throughout space in this case. As a result, $\rho^*$ must be an equilibrium density having the Gibbs form
\be
\rho^*(x) =c\, e^{-2 U(x)/\sigma^2},
\label{eqgibbs1}
\ee
where $U(x)$ is the potential associated with the force by $F(x) = -U'(x)$ and $c$ is a normalization constant.

\begin{figure}[t]
\centering
\includegraphics{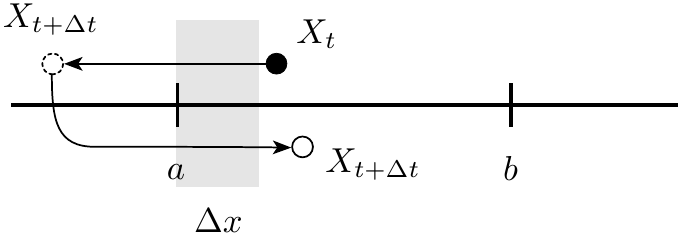}
\caption{Mechanical reflection of the Langevin dynamics at a boundary. If the state $X_t$ crosses the boundary $a$ during one time step, then it is reflected back to $[a,b]$ in a mirror-like way with respect to $a$. A similar reflection is applied to the boundary at $b$. This reflection rule assumes that the reflected point still falls within $[a,b]$, which is the case if $\Delta t$ is sufficiently small. The boundary layer is shown in grey.}
\label{figreflect1}
\end{figure}

Perfect reflections at the boundaries can also be imposed directly at the level of $X_t$ by adding to the SDE a new ``noise'' term given by the increment of the ``local time'' at the boundaries, which essentially represents the amount of time that $X_t$ spends near $a$ or $b$ \cite{schuss2013}. This mathematical construction, due to Skorokhod \cite{skorokhod1961}, provides a rigorous way to study reflections in diffusions, but will not be used here as it is too abstract for our purposes. Instead, we think of $X_t$ as evolving on $\reals$ in discrete time according to the Euler--Maruyama scheme
\be
X_{t+\Delta t} =X_t +F(X_t)\Delta t +\sigma \sqrt{\Delta t} Z,
\ee
and we simply reflect the update $X_{t+\Delta t}$ back inside $[a,b]$ whenever it falls outside this interval \cite{schuss2013}, as illustrated in Fig.~\ref{figreflect1}. Here, $\Delta t$ is the discretized time step while $Z\sim N(0,1)$ is a standard normal random variable. 

The typical distance $\Delta x$ from a boundary within which $X_t$ can cross it in a single time step defines an exclusion zone, called the \emph{boundary layer,} which represents an artefact or error of the Euler--Maruyama scheme. A similar boundary layer is found if one simulates the reflections in a ``soft'' way by adding a fictitious potential to $U(x)$ to create strong repulsive walls at $a$ and $b$ \cite{menaldi1983,pettersson1997,kanagawa2000}. In both cases, the thickness of the layer generally decreases to $0$ as $\Delta t\ra 0$, so they give a good approximation of the dynamics of $X_t$ when $\Delta t$ is sufficiently small. In this limit, it should also be clear that a ``particle'' with state $X_t$ entering the boundary layer will leave it instantaneously, so that the net number of crossings at the layer is 0. Viewing $\rho(x,t)$ as representing the density of an ensemble of such particles and $J_{F,\rho}(x,t)$ as their flux, we then recover the zero-current condition \eqref{eqcurr1} when the layer disappears.

\section{Markov operators}
\label{secmop}

For the results to come, it is important to rewrite the Fokker--Planck equation in operator form as
\be
\p_t \rho(x,t) = \cL^\dag \rho(x,t)
\ee
in order to identify the Fokker--Planck operator
\be
\cL^\dag = -\p_xF +\frac{\sigma^2}{2}\p_{xx}
\label{FPop}
\ee
as the generator of the time evolution of $\rho(x,t)$, starting from some initial density $\rho(x,0)$. The domain $\cD(\cL^\dag)$ of this linear operator is naturally the set of all probability densities that (i) can be normalized on $[a,b]$; (ii) satisfy the zero-current condition \eqref{eqcurr1}, which has the form of a Robin (mixed) boundary condition on $\rho$; and (iii) are twice-differentiable, since $\cL^\dag$ is a second-order differential operator. 

Dual to $\cL^\dag$ is the Markov generator
\be
\cL = F\p_x+\frac{\sigma^2}{2}\p_{xx}
\label{generator}
\ee 
which governs the evolution of expectations according to
\be
\p_t E[g(X_t)]=E[\cL g(X_t)],
\ee
where $g$ is a test function and $E[\cdot]$ denotes the expectation with respect to $\rho(x,t)$. The two generators are dual or adjoint to each other in the sense that
\be
\langle \cL^\dag \rho,g\rangle = \langle \rho, \cL g\rangle
\ee
with respect to the standard inner product used in the theory of Markov processes, namely,
\be
\langle \rho,g\rangle = \int_a^b \rho(x)g(x)\, dx = E[g(X)],
\label{eqip1}
\ee
where $\rho$ is an arbitrary density in $\cD(\cL^\dag)$ and $g$ is a test function. From this definition, as well as the expressions (\ref{FPop}) and (\ref{generator}) for $\cL^{\dagger}$ and $\cL$, we find 
\be
\langle \cL^\dag \rho,g \rangle =\langle \rho, \cL g \rangle -g(x) J_{F,\rho}(x)\big|_a^b - \frac{\sigma^2}{2}\rho(x) g'(x)\big|_a^b 
\label{integrationbyparts}
\ee
via integration by parts. Given that the current $J_{F,\rho}$ vanishes at the boundaries, we must then require that the test functions $g$ acted on by $\cL$ satisfy 
\be
g'(a^+) = 0=g'(b^-) 
\label{eqmarkovcond1}
\ee
in order for the boundary term in \eqref{integrationbyparts} to vanish and, thus, for the operators $\cL$ and $\mathcal{L^{\dagger}}$ to be proper duals defined independently of any specific $\rho$ or $g$ \cite{nagasawa1961}. From this result, the domain $\cD(\cL)$ of $\cL$ is then defined to be the set of test functions that (i) have finite expectation (finite inner product); (ii) satisfy the zero-derivative condition \eqref{eqmarkovcond1}, which is a Neumann boundary condition; and (iii) are twice differentiable.

\section{Large deviations}

We now come to the main point of our work, namely, to understand how reflecting boundaries determine the large deviations of dynamical observables. To this end, we briefly recall the definition of the rate function, which characterises the probability distribution of $S_T$ in the long-time limit, and present the spectral problem underlying the calculation of this function. We then explain how the boundary conditions normally applied to this spectral problem in the case of unbounded diffusions must be modified to account for perfect reflections, and how these new boundary conditions affect the properties of the driven process, which explains how fluctuations of $S_T$ arise in time. 

\subsection{Large deviation functions}

We consider an ergodic reflected diffusion $X_t$, as defined in the previous section, and a dynamical observable $S_T$ of this process having the form \eqref{eqobs1}. We are interested in finding the \emph{rate function} of $S_T$, defined by the limit
\be
I(s) = \lim_{T\ra\infty} -\frac{1}{T}\ln P_T(s),
\ee
where $P_T(s)$ is the probability distribution of $S_T$. In practice, it is very difficult to obtain this distribution exactly, which is why the rate function is sought instead. Indeed, for a large class of Markov processes and observables, it can be shown that 
$
P_T(s)\approx e^{-T I(s)}
$
with sub-exponential corrections in $T$, so that $I(s)$ effectively describes the shape of $P_T(s)$ at leading order in $T$ \cite{touchette2009}. This holds in the limit of large integration times $T$, typically much longer than the relaxation time-scale of $X_t$.
 
To calculate the rate function, we use the fact that it is dual to another large deviation function, called the \emph{scaled cumulant generating function} (SCGF) \cite{dembo1998} and defined as
\be
\lambda(k) = \lim_{T\ra\infty} \frac{1}{T}\ln E[e^{Tk S_T}],\quad k\in\reals.
\ee
Using the Feynman--Kac formula, it can be shown \cite{touchette2017} that this function coincides with the dominant (real) eigenvalue of a linear operator, called the \emph{tilted generator}, having the form 
\be
\cL_k= \cL + k f,
\label{tilted}
\ee
where $\cL$ is the generator of $X_t$ and $f$ is the function appearing in the definition \eqref{eqobs1} of the observable $S_T$. Having the SCGF, we then obtain $I(s)$ by taking a Legendre transform, so that
\be
I(s) = k(s) s-\lambda(k(s)),
\label{eqlf1}
\ee
where $k(s)$ is the unique root of $\lambda'(k) = s$. This essentially holds if $\lambda(k)$ is differentiable and strictly convex.

The calculation of the rate function thus reduces to solving the spectral problem
\be
\cL_k r_k(x) = \lambda(k) r_k(x),
\label{eqse1}
\ee
where $r_k$ is the eigenfunction of $\cL_k$ corresponding to the dominant eigenvalue $\lambda(k)$. Given that $\cL_k$ is not in general Hermitian, this spectral problem must be solved in conjunction with its dual 
\be
\cL_k^{\dag} l_k(x) = \lambda(k) l_k(x),
\label{eqse2}
\ee
where $l_k$ is the eigenfunction of $\cL_k^\dag=\cL^\dag+kf$ with the same dominant eigenvalue $\lambda(k)$. The boundary conditions that we must impose on these two spectral equations to obtain $\lambda(k)$ are discussed next. Independently of these conditions, $r_k$ and $l_k$ are dual functions with respect to the standard inner product \eqref{eqip1}, so they must satisfy $\langle l_k, r_k\rangle <\infty$. They are also positive functions, since they are the dominant modes of $\cL_k$ and $\cL_k^\dag$, respectively. In the literature \cite{chetrite2014}, it is common to normalize them in such a way that
\be
\int_a^b l_k(x)r_k(x) \, dx =\langle l_k,r_k\rangle= 1
\label{normalization1}
\ee
and 
\be
\int_a^b l_k(x) \, dx =\langle l_k,1\rangle= 1.
\label{normalization2}
\ee
The latter integral is consistent, as we will see, with the fact that $\cL_k^\dag$ is related to the Fokker--Planck operator acting on normalized densities.

\subsection{Boundary conditions}

The boundary conditions that must be imposed on $r_k$ and $l_k$ to solve the spectral equations \eqref{eqse1} and \eqref{eqse2} are determined by considering, as done before, the boundary term that arises when transforming $\cL_k$ to its dual $\cL_k^\dag$. Since these operators differ from $\cL$ and $\cL^\dag$, respectively, only by the constant term $kf$, which produces no boundary terms of its own, the result of \eqref{integrationbyparts} can readily be used to obtain
\be
\langle \cL_k^\dag l_k, r_k\rangle =\langle l_k, \cL_k r_k\rangle- r_k(x)J_{F,l_k}(x)\big|_a^b - \frac{\sigma^2}{2}l_k(x)r'_k(x)\big|_a^b,
\label{eqbdt1}
\ee
where
\be
J_{F,l_k}(x)=F(x)l_k(x)-\frac{\sigma^2}{2}l'_k(x)
\label{eqlcurr1}
\ee
is the probability current associated with $l_k$.

For diffusions evolving in $\reals$ without boundaries, we see that the boundary term in \eqref{eqbdt1} vanishes for any $r_k(x)$ and $l_k(x)$ if the latter eigenfunction decays to 0 sufficiently fast as $|x|\ra\infty$. This is consistent with the normalization integral in \eqref{normalization2} and implies with $l_k>0$ that both $l'_k(x)$ and $J_{F,l_k}(x)$ vanish as $|x|\ra\infty$. There is no condition on $r_k$ alone, as such, since the Markov generator $\cL$, and by extension $\cL_k$, have no natural boundary conditions, except for the fact that both act on functions that have finite inner product, which for $\cL_k$ translates into the integral condition in \eqref{normalization1}, normalized to 1.

For reflected diffusions, the normalization integrals \eqref{normalization1} and \eqref{normalization2} are no longer sufficient on their own to define boundary conditions on $r_k$ and $l_k$. Instead, we must impose
\be
J_{F,l_k}(a^+) = 0 = J_{F,l_k}(b^-)
\label{boundaryl}
\ee
and 
\be
r_k'(a^+) = 0 = r_k'(b^-)
\label{boundaryr}
\ee
in order for the boundary term in \eqref{eqbdt1} to vanish and, importantly, for $\cL_k^\dag$ and $\cL_k$ to have boundary conditions that are consistent with those imposed on $\cL^\dag=\cL^\dag_{k=0}$ and $\cL=\cL_{k=0}$, respectively, when $k=0$. The same conditions also follow by noticing again that the boundary term in the duality of $\cL_k^\dag$ and $\cL_k$ does not depend on $kf$, so that \eqref{boundaryl} simply extends the zero-current condition of $\cL^\dag$ to $\cL_k^\dag$, while \eqref{boundaryr} extends the Neumann boundary condition of $\cL$ to $\cL_k$. As a result, $\cD(\cL_k^\dag)=\cD(\cL^\dag)$ and $\cD(\cL_k)=\cD(\cL)$. These boundary conditions must be imposed, incidentally, not just on the eigenfunctions related to the dominant eigenvalue, but to all eigenfunctions, thereby determining the full spectrum of $\cL_k$ which is conjugate to the spectrum of $\cL_k^\dag$. For large deviation calculations, however, we only need the dominant eigenvalue, which is real, and the corresponding eigenfunctions, which are positive.

\subsection{Driven process}

While the SCGF and the rate function describe the fluctuations of $S_T$, they do not provide any information about how these fluctuations are created in time. Recently, it has been shown \cite{chetrite2013,chetrite2014,chetrite2015} that this information is provided by a modified Markov process $\hX_t$, called the \emph{effective} or \emph{driven process}, which represents in some approximate way the original diffusion $X_t$ conditioned on the fluctuation $S_T=s$ and so describes the paths of $X_t$ that lead to or create that fluctuation. 

The details of this process can be found in many studies \cite{chetrite2013,chetrite2014,chetrite2015}, which provide a full derivation of its properties and interpretation. In the context of ergodic diffusions, the driven process satisfies the new SDE
\be
d\hX_t  = F_k(\hX_t)dt +\sigma dW_t,
\ee
where the \emph{driven force} $F_k$ is a modification of the force $F$ given in terms of the dominant eigenfunction $r_k$ by
\be
F_k(x) = F(x) +\sigma^2 \frac{r_k'(x)}{r_k(x)}.
\label{eqddf1}
\ee
The parameter $k$ is the same as that entering in the SCGF: its value is set for a given fluctuation $S_T=s$ according to the Legendre transform \eqref{eqlf1} as the root of $\lambda'(k)=s$ or, equivalently, as $k=I'(s)$ \footnote{These two relations, which follow from the Legendre transform \eqref{eqlf1}, hold when $I(a)$ is strictly convex \cite{chetrite2014}.}. With this choice, the stationary density of the driven process, which is known to be given \cite[Sec.~5.4]{chetrite2014}  by
\be
\rho_k^*(x) = r_k(x)l_k(x),
\label{eqstatdistdrive1}
\ee
is such that
\be
\int_a^b \rho_k^*(x) f(x)\, dx = s.
\label{eqdrivenobs1}
\ee
This shows that we can also interpret the driven process as a change of process that transforms the fluctuation $S_T=s$ into a typical value realized by $\hX_t$ in the ergodic limit. The change of drift and density also modifies the stationary current to
\be
J_{F_k,\rho_k^*}(x) = F_k(x)\rho_k^*(x) -\frac{\sigma^2}{2} \rho_k^*(x)'.
\ee
Note that for $k=0$, $r_{k=0}=1$ while $l_{k=0}=\rho^*$, leading to $F_{k=0}=F$, $\rho_{k=0}^*=\rho^*$ and $J_{F_{k=0},\rho^*_{k=0}}=J_{F,\rho^*}$. 

For an ergodic diffusion $X_t$ evolving on $[a,b]$ with reflecting boundaries, the driven process $\hX_t$ also evolves on $[a,b]$, since it represents a conditioning of $X_t$, and inherits for the same reason a zero-current boundary conditions at $x=a^+$ and $x=b^-$, given in terms of the driven current by
\be
J_{F_k,\rho_k^*}(a^+)=0=J_{F_k,\rho_k^*}(b^-).
\label{eqcurrdrive1}
\ee
This can be verified, in fact, independently of the interpretation of $\hX_t$ by applying the boundary conditions on $r_k$ and $l_k$ discussed previously to the driven process. First, note that the Neumann boundary conditions \eqref{boundaryr} on $r_k$ implies with the definition of the driven force \eqref{eqddf1} that the latter is constrained to satisfy
\be
F_k(a^+)=F(a^+)\quad\textrm{and}\quad F_k(b^-)=F(b^-).
\label{eqeqdrift1}
\ee
Hence, the drift at the boundaries is not modified at the level of the driven process, although it is modified in the interior of $[a,b]$, as we will see in the next section with specific examples. This is a significant result of our work, which also implies that any density $\rho$ satisfying a zero-current condition at the boundaries with respect to the original drift must also satisfy a zero-current condition at the boundaries with respect to the driven force. In other words, $J_{F,\rho}(a^+)=0$ is equivalent to $J_{F_k,\rho}(a^+)=0$ and similarly for $x=b^-$.

Next, we note that the boundary conditions (\ref{boundaryl}) and (\ref{boundaryr}) can be combined as 
\be
r_k(x)J_{F,l_k}(x)\big|_{x = a^+,b^-} = 0
\label{boundaryterm1}
\ee
and 
\be
\frac{\sigma^2}{2} l_k(x) r_k'(x)\big|_{x = a^+,b^-} = 0.
\label{boundaryterm2}
\ee
We recognize these as the boundary terms in (\ref{eqbdt1}), which can be combined to obtain
\be
\left[ r_k(x) J_{F,l_k} (x) - \frac{\sigma^2}{2} l_k(x) r_k'(x) \right]_{x = a^+,b^-} = 0.
\ee
Upon expanding the expression of the current associated with $l_k$ and combining terms, we then obtain 
\be 
\left[ F(x) r_k(x) l_k(x) - \frac{\sigma^2}{2} \left(r_k(x) l_k(x)\right)' \right]_{x = a^+,b^-} = 0,
\label{zerocurrentpk}
\ee
which is a zero-current condition for the product $r_k l_k$. From this result, we finally recover \eqref{eqcurrdrive1} using \eqref{eqeqdrift1} and the fact that $r_kl_k$ is the stationary density of the driven process, as shown in \eqref{eqstatdistdrive1}. 

This confirms in a more direct way that the driven process also evolves in $[a,b]$ with perfect reflections at the boundaries. Of course, since we are dealing with one-dimensional diffusions, the stationary current of the driven process must vanish not just at the boundaries but over the whole of $[a,b]$, similarly to the original diffusion. This is a known result in the theory of the driven process: if the original Markov process is reversible, then so is the driven process in the case where the dynamical observable has the form of \eqref{eqobs1} \cite{chetrite2014}. This does not mean that $\rho^*$ and $\rho^*_k$ are the same in $[a,b]$ -- we will see illustrations of this point in the next section. Moreover, note that requiring $J_{F_k,\rho_k}=0$ at the boundaries is not sufficient to define boundary conditions for $r_k$ and $l_k$ since the driven current mixes both $F_k$ and $\rho_k$. These conditions are determined again by studying the boundary term arising in the duality between $\cL_k$ and $\cL_k^\dag$.

\subsection{Symmetrization}

The spectral calculation of the SCGF is complicated by the fact that the tilted generator $\cL_k$ is not Hermitian in general. A significant simplification is possible when the spectrum of $\cL_k$ is real, as is the case, for example, when $X_t$ is an ergodic, gradient diffusion characterized by the Gibbs stationary distribution \eqref{eqgibbs1} and $S_T$ is defined as in \eqref{eqobs1}. Then it is known \cite{touchette2017} that $\cL_k$ can be transformed in a unitary way, therefore preserving its spectrum, to the following Hermitian operator:
\be 
\cH_k =\sqrt{\rho^*}\cL_k \frac{1}{\sqrt{\rho^*}}= \frac{\sigma^2}{2} \p_{xx} - V_k,
\label{hermitian}
\ee
which involves a quantum-like potential given by 
\be 
V_k(x) = \frac{|U'(x)|^2}{2\sigma^2} - \frac{U''(x)}{2} - kf,
\label{drivenpot}
\ee
where $U$ is the potential of the gradient force and $f$ the function defining the observable $S_T$.

Since the new operator $\cH_k$ has the same spectrum as $\cL_k$, the spectral problem associated with the SCGF becomes
\be
\cH_k \psi_k = \lambda(k)\psi_k,
\label{eqspecprob1}
\ee
where $\psi_k$ is the corresponding eigenfunction related to $r_k$ and $l_k$ by 
\be
\psi_k(x) = \sqrt{\rho^*(x)}r_k(x) = l_k(x)/\sqrt{\rho^*(x)}.
\label{psirelate}
\ee
From \eqref{normalization1}, $\psi_k$ thus satisfies the normalization condition
\be
\int_a^b \psi_k(x)^2\, dx=1,
\label{eqpsinorm1}
\ee
similarly to that found in quantum mechanics, but for a real eigenfunction. Moreover, using $\psi_k^2=r_kl_k=\rho^*_k$ we find with \eqref{zerocurrentpk} that $\psi_k$ must satisfy for reflected diffusions the zero-current boundary conditions
\be
J_{F,\psi_k^2}(a^+)=0=J_{F,\psi_k^2}(b^-).
\ee
This can be expressed more explicitly as
\be
\psi_k'(a^+) = \frac{F(a^+)}{\sigma^2} \psi_k(a^+),
\label{boundarypsi}
\ee
with a similar expression holding for $x=b^-$.

\section{Examples}
\label{secex}

We illustrate in this section the results derived before by applying them to two simple reflected diffusions obtained by constraining the Brownian motion with negative drift and the Ornstein--Uhlenbeck process on the half line. Each of these diffusions can be solved exactly and give rise to interesting properties for the driven process in the presence of a reflecting boundary. Further applications for diffusions evolving in higher dimensions are mentioned in the conclusion.

\subsection{Reflected Ornstein--Uhlenbeck process}

The first example that we consider is the reflected Ornstein--Uhlenbeck process (ROUP) satisfying the SDE 
\be
dX_t = - \gamma X_t  dt + \sigma dW_t
\ee
with $\gamma>0$, $\sigma>0$, $X_t \in [0,\infty)$, and perfect reflection at $x = 0$. This diffusion is used in engineering as a continuous-space model of queues in the heavy-traffic regime \cite{giorno1986,ward2003,linetsky2005} and represents, more physically, the dynamics of a Brownian particle attracted to a reflecting wall by a linear force induced, for example, by laser tweezers \cite{ashkin1997} or an ac trap \cite{cohen2005b}. For this process, we consider the observable
\be
S_T = \frac{1}{T} \int_0^T X_t \, dt,
\label{eqobs2}
\ee
which for laser tweezers is related to the mechanical power expended by the lasers on the particle \cite{zon2003a}. 

The tilted generator $\cL_k$ associated with this process and observable is given from (\ref{generator}) and (\ref{tilted}) by 
\be
\cL_k = -\gamma x \p_x + \frac{\sigma^2}{2} \p_{xx} + k x
\ee
and is clearly non-Hermitian. However, because the ROUP is gradient and ergodic on $[0,\infty)$, we can apply the symmetrization transform described before by substituting the potential $U(x) = \gamma x^2/2$ in (\ref{drivenpot}) to obtain from \eqref{hermitian}
\be
\cH_k = \frac{\sigma^2}{2} \p_{xx} - \frac{\gamma^2 x^2}{2 \sigma^2} + \frac{\gamma}{2} + kx.
\ee
This defines with \eqref{eqspecprob1} the spectral problem that we need to solve in order to obtain the SCGF. The boundary condition (\ref{boundarypsi}) reduces in this case to the simple Neumann condition
\be 
\psi_k'(0) = 0.
\label{boundaryROUP}
\ee
For $x=\infty$, the normalization \eqref{eqpsinorm1} requires that $\psi_k(x)^2\ra 0$ as $x\ra\infty$ and, therefore, $\psi_k(x)\ra 0$ as $x\ra\infty$.

\begin{figure}[t]
\centering
\includegraphics{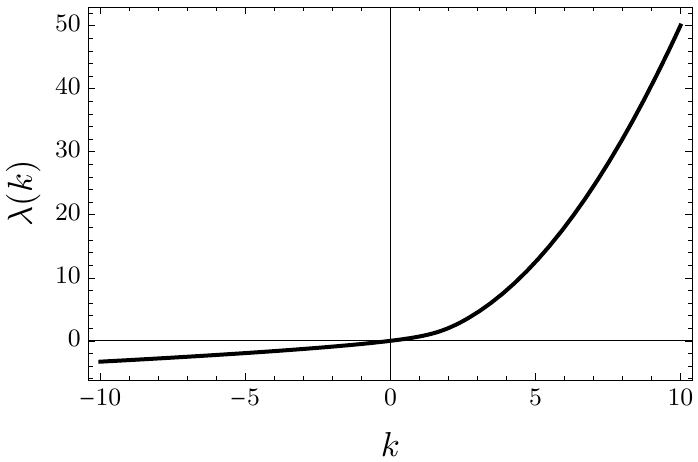}%
\caption{SCGF $\lambda(k)$ for the ROUP with linear observable. Parameters: $\sigma=1$ and $\gamma=1$.} 
\label{fig:ROUPSCGF}
\end{figure}

The spectral problem \eqref{eqspecprob1} defines for $\cH_k$ above a differential equation whose solutions are taken from the class of parabolic cylinder functions $D_\nu(z)$, the dominant solution $\psi_k$ having the form
\be 
\psi_k(x) = D_{\xi(k)} \left(\frac{\sqrt{2\gamma} x}{\sigma} - \frac{\sqrt{2} k \sigma}{\gamma^{3/2}} \right)
\label{cylinder}
\ee
up to a normalization constant, where we have defined 
\be 
\xi(k) = \frac{k^2 \sigma^2 - 2 \gamma^2 \lambda(k)}{2\gamma^3}.
\ee
This solution decays to 0 at infinity. By imposing the boundary condition (\ref{boundaryROUP}) and using well-known properties of the parabolic cylinder functions, we find that $\lambda(k)$ is determined implicitly by the transcendental equation 
\be
\frac{k \sigma}{\sqrt{2} \gamma^{3/2}} D_{\xi(k)} \left(- \frac{\sqrt{2} k \sigma}{\gamma^{3/2}} \right) + D_{\xi(k) + 1} \left(\frac{\sqrt{2} k \sigma}{\gamma^{3/2}} \right) = 0.
\label{eqtranseq1}
\ee
To be more precise, $\lambda(k)$ is the largest root of this equation; the other roots, which are all real, give the rest of the spectrum of $\cH_k$ and therefore of $\cL_k$. 

\begin{figure}[t]
\centering
\includegraphics{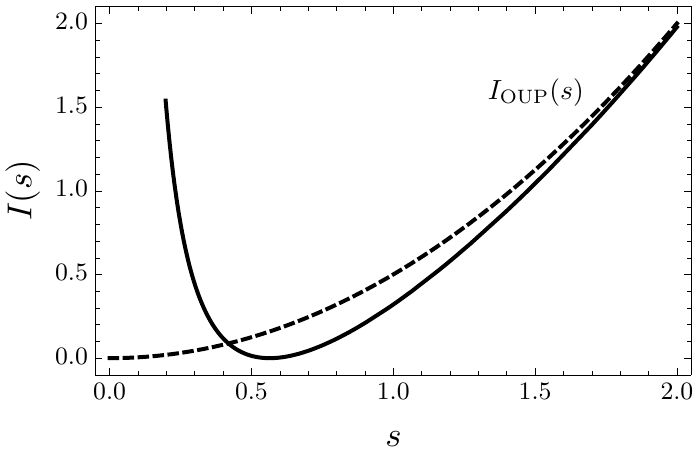}
\caption{Rate function $I(s)$ for the ROUP with linear observable, compared with the rate function of the normal OUP. Parameters: $\sigma=1$ and $\gamma=1$.} 
\label{fig:ROUPrate}
\end{figure}

We show in Fig.~\ref{fig:ROUPSCGF} the plot of $\lambda(k)$ obtained by solving the transcendental equation numerically for a given set of parameters $\gamma$ and $\sigma$ and different values of $k$. We can see that $\lambda(0)=0$, as follows from the definition of the SCGF, and that $\lambda(k)$ appears to be differentiable and strictly convex. As a result, we can take the Legendre transform \eqref{eqlf1} to obtain the rate function $I(s)$, shown in Fig.~\ref{fig:ROUPrate}. There we see that $I(s)$ has two very different branches on either side of the minimum and zero $s^*$, corresponding to the most probable value of $S_T$ at which $P_T(s)$ concentrates exponentially as $T\ra\infty$. The left branch, related to the $k<0$ branch of the SCGF, is steep and therefore indicates that small fluctuations of $S_T$ close to $0$, produced by paths that stay close to the reflecting boundary, are very unlikely. The right branch, on the other hand, is less steep and has overall the shape of a parabola, signalling that the large fluctuations of $S_T$ are Gaussian-distributed. 

This is confirmed by comparing $I(s)$ with the rate function of $S_T$ for the \emph{normal} Ornstein--Uhlenbeck process (OUP) evolving on the whole of $\reals$, which is known to be
\be
I_\oup(s)=\frac{\gamma^2 s^2}{2\sigma^2}.
\ee
This rate function is shown with the dashed line in Fig.~\ref{fig:ROUPrate} and closely approximates, as can be seen, the upper tail of the rate function obtained with reflection. This is explained by noting that large fluctuations of $S_T$ arise from paths that venture far from the reflecting boundary, and so do not ``feel'' its influence. The zero at $s^*$ is itself a product of the boundary, since $s^*=0$ in the normal OUP. We can determine its value by noting from the ergodic theorem that the most probably value of $S_T$ is the stationary expectation
\be
s^*=\int_0^\infty x\rho^*(x)\, dx.
\label{eqcstar1}
\ee
Here $\rho^*(x)$ is a truncated Gaussian density with potential $U(x)=\gamma x^2/2$ restricted to $x\in [0,\infty)$, leading in the integral above to $s^*=\sigma/\sqrt{\pi \gamma}$, which gives $s^*=1/\sqrt{\pi}\approx 0.564$ for the parameters used in Fig.~\ref{fig:ROUPrate}.

\begin{figure*}[t]
\centering
\includegraphics{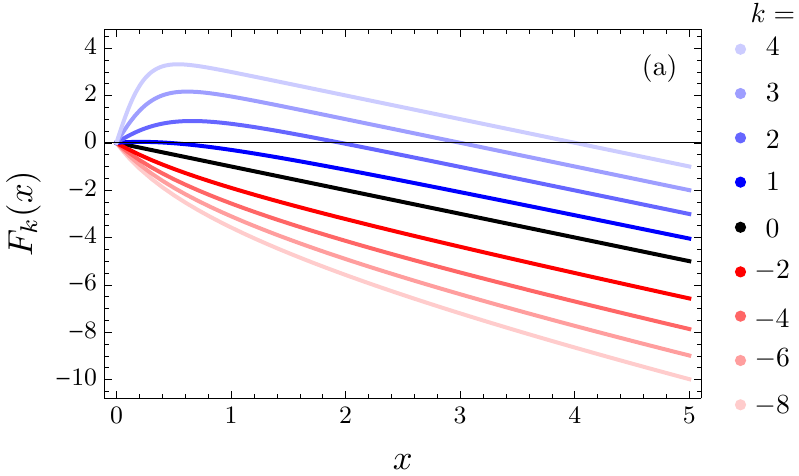}%
\hspace*{0.3in}%
\includegraphics{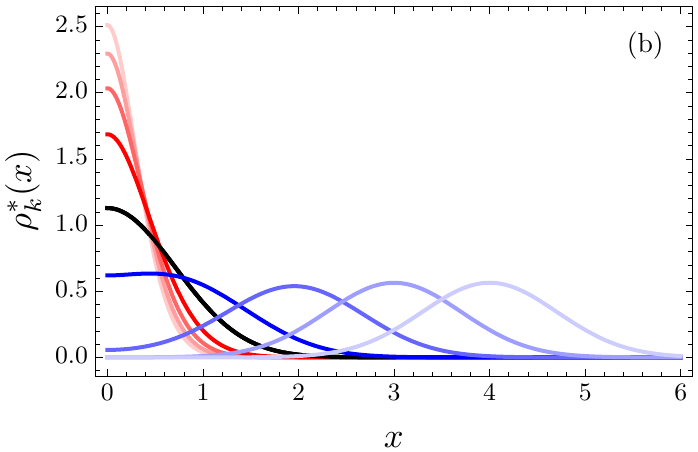}
\caption{(Color online) (a) Drift $F_k(x)$ of the driven process for the ROUP with linear observable plotted for different values of $k$ (see legend). (b) Corresponding stationary density $\rho_k^*(x)$. Parameters: $\sigma=1$ and $\gamma=1$.} 
\label{fig:ROUPdrivenf}
\end{figure*}

The large Gaussian fluctuations of $S_T$ are also confirmed by analyzing the driven force $F_k$, which we find from \eqref{eqddf1} using the expression \eqref{cylinder} for $\psi_k$ and the relation \eqref{psirelate}, yielding
\be
F_k(x) = \sqrt{2}\gamma\sigma\frac{\frac{1}{2}\eta D_{\xi(k)}(\eta)- D_{\xi(k)+1}(\eta)}{D_{\xi(k)}(\eta)},
\label{eqdriftroup1}
\ee
where
\be
\eta = \frac{\sqrt{2\gamma} x}{\sigma} - \frac{\sqrt{2} k \sigma}{\gamma^{3/2}}.
\ee
To obtain this expression, we have also used some identities of the parabolic cylinder functions \cite{buisson2020}. The result is plotted in Fig.~\ref{fig:ROUPdrivenf}(a) for different values of $k$, together with the properly normalized $\rho_k^*(x) = \psi_k(x)^2$ in Fig.~\ref{fig:ROUPdrivenf}(b).

Comparing the two plots, we see that $F_k(x)$ has two zeros for $k>0$ and, therefore, two critical points: one at $x=0$, which gives rise to a local minimum in $\rho_k^*(x)$, and another at some value $x>0$, which is responsible for the maximum of $\rho_k^*(x)$. Around the latter critical point, $F_k(x)$ is approximately linear with slope $-\gamma$, implying that $\rho_k^*(x)$ is approximately Gaussian around its maximum. Moreover, as $k$ is increased, we see that the maximum of $\rho_k^*(x)$ moves away from $x=0$, showing that the driven process is repelled from the boundary, thus creating larger typical values of $S_T$, similarly to the driven process of the normal OUP \cite{chetrite2014}. The difference between the two processes is that the driven force of the ROUP is ``bent'' near the boundary so as to have $F_k(0)=F(0)$, consistently with \eqref{eqeqdrift1}, whereas that of the OUP is always linear \cite{chetrite2014}.

For $k<0$, the picture is different. The driven force $F_k(x)$ only has a single critical point at $x=0$, creating the maximum of $\rho_k^*(x)$ seen in Fig.~\ref{fig:ROUPdrivenf}(b). As $k\ra-\infty$, $\rho_k^*(x)$ gets more concentrated at $x=0$, as the driven process is attracted to the reflecting boundaries, creating smaller fluctuations of $S_T$. Such a behavior is very unlikely in the ROUP, which is why the rate function is steep close to $s=0$. In fact, it is steeper than a parabola because $F_k(x)$ is not linear away from $x=0$: its curvature becomes more pronounced for large negative values of $k$, leading $\rho_k^*(x)$ to be non-Gaussian. Note in all cases that $F_k(0)=F(0)=0$, consistently again with \eqref{eqeqdrift1}.

This analysis of $F_k(x)$ is useful not only in understanding the different stochastic mechanisms underlying or ``producing'' different fluctuations of $S_T$, but also in deriving accurate approximations of the rate function by following three steps \cite{chetrite2015}. First, approximate $F_k$ by some function, say $\tF_\theta$, where $\theta$ denotes a set of parameters. Second, calculate the stationary distribution $\trho^*_\theta$ that results from this approximation. Third and finally, calculate the ergodic expectation of $S_T$ with respect to $\trho^*_\theta$:
\be
s_\theta = \int_a^b f(x) \trho_\theta^*(x)\, dx,
\ee
as well as the integral
\be
K_\theta = \frac{1}{2\sigma^2}\int_a^b [F(x)-\tF_\theta(x)]^2 \trho_\theta^*(x)\, dx.
\label{eqcost1}
\ee
Then
\be
I(s_\theta) \leq K_\theta
\ee
with equality if and only if $\tF_\theta=F_k$ at $s_\theta$ \cite{chetrite2015}.

It is beyond this paper to prove this result (see \cite{chetrite2015}). At this point we only want to use it to find useful upper bounds on $I(s)$, beginning with the left branch of this function, related as mentioned before to the $k<0$ branch of $\lambda(k)$. In this case, we know that $F_k(x)$ has a single critical point at $x=0$, so we approximate it in a simple way as
\be
\tF_\theta(x) = -\theta x.
\ee
Only $\theta\geq \gamma$ need be considered, since it is clear from Fig.~\ref{fig:ROUPdrivenf}(b) that $F'_k(0)\leq -\gamma$ for $k\leq 0$. This approximation retains the linear form of $F(x)$, which means that $\trho^*_\theta(x)$ is the same truncated Gaussian density as the ROUP but with $\gamma$ replaced by $\theta$. As a result, we have $s_\theta = \sigma/\sqrt{\pi \theta}$ and obtain
\be
K_\theta = \frac{(\theta-\gamma)^2}{4\theta}
\ee
by direct integration of \eqref{eqcost1}. Changing the $\theta$ variable to $s$ with $s_\theta=s$, we then find 
\be
\tI(s) = K_{\theta(s)} = \frac{\pi}{4}\left(\frac{\sigma}{\pi s}-\frac{\gamma s}{\sigma}\right)^2
\label{eqapproxrate1}
\ee
as our approximation of $I(s)$ for $s\in (0,s^*]$.

We do not compare this result with the exact rate function obtained from the spectral calculation, as the two are nearly indistinguishable \cite{buisson2020}. They agree exactly at $s^*$, since $\theta=\gamma$ recovers the drift of the ROUP, and start to differ only close to $s=0$ because $F_k$ is curved there, as we noticed, whereas $\tF_\theta$ is not. Note that the divergence of $I(s)$ near $s=0$ predicted by the approximation above is $\tI(s)\sim \sigma^2/(4\pi s^2)$ as $s\ra 0$, which is independent of $\gamma$.

Similar calculations can be carried out for $k>0$ to approximate $I(s)$ for $s>s^*$. In that case, the form of $F_k$ suggests that we use 
\be
\tF_\theta(x) = -\gamma x+\theta
\label{eqansatzforce2}
\ee
as an approximate drift parameterized by $\theta>0$. The associated stationary density $\trho_\theta^*$ can also be obtained in closed form and yields, after solving a number of Gaussian integrals \cite{buisson2020},
\be
s_\theta = \sqrt{\frac{\sigma^2}{\pi\gamma}} \frac{e^{-\theta^2/(\gamma \sigma^2)}}{1+\erf[\theta/(\sqrt{\gamma}\sigma)]}+\frac{\theta}{\gamma} 
\label{eqstypdriven1}
\ee
and
\be
K_\theta = \frac{\theta^2}{2\sigma^2}.
\ee
The presence of the error function in $s_\theta$ prevents us from expressing $K_\theta$ in closed form as a function of $s$, as in \eqref{eqapproxrate1}. However, the result clearly shows that the rate function becomes a parabola as $\theta\ra \infty$, for in that limit, $s_\theta\sim \theta/\gamma$, leading to  $\tI(s) = K_{\theta(s)}=I_\oup(s)$ as $s\ra\infty$.

\subsection{Reflected Brownian motion with drift}

The second example we consider is the reflected Brownian motion with drift (RBMD) governed by the SDE 
\be
dX_t = -\mu dt + \sigma dW_t,
\ee
where $\mu>0$, $\sigma>0$, and $X_t \in [0,\infty)$, with reflection imposed at $x = 0$ \cite{linetsky2005}. This process was studied by Fatalov \cite{fatalov2017} and models, similarly to the ROUP, the dynamics of a particle attracted by a force to a reflecting wall. The force is now constant and can be viewed, for instance, as the gravity pulling vertically on a Brownian particle in a container. As for the ROUP, we consider the linear observable defined in \eqref{eqobs2}.

The tilted generator associated with the large deviations of $S_T$ for this process is given by \eqref{generator} with $F(x)=-\mu$ and leads, after symmetrization with the corresponding potential $U(x) = \mu x$, to the Hermitian operator
\be
\cH_k = \frac{\sigma^2}{2} \p_{xx} - \frac{\mu^2}{2 \sigma^2} + kx.
\label{eqham2}
\ee
This defines with \eqref{eqspecprob1} the spectral problem that gives the SCGF, which needs to be solved with the Robin boundary condition (\ref{boundarypsi})
\be 
\psi_k'(0) = -\frac{\mu}{\sigma^2} \psi_k(0)
\label{RBMDboundary}
\ee
at $x = 0$ and $\psi_k(x)\ra 0$ as $x\ra\infty$. 

\begin{figure}[t]
\centering
\includegraphics{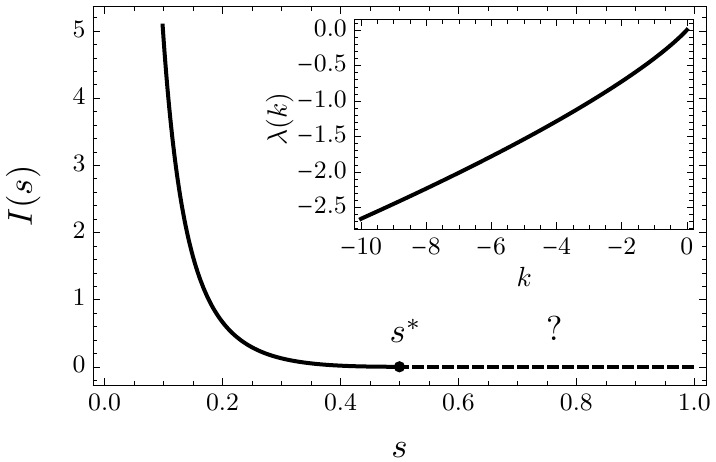}
\caption{SCGF $\lambda(k)$ for the RBMD with linear observable (inset) and corresponding rate function $I(s)$. Parameters: $\sigma=1$ and $\mu=1$. The rate function is defined by spectral means for $0<s\leq s^*=\sigma^2/(2\mu)$; it is unknown above $s^*$, as indicated by the question mark.}
\label{fig:RBMDLD}
\end{figure}

The eigenfunctions of $\cH_k$ satisfying these conditions are now expressed in terms of Airy functions of the first kind. The dominant eigenfunction is
\be 
\psi_k(x) = \Ai \left[\left(-\frac{\sigma^2}{2k}\right)^{2/3} \left(-\frac{2kx}{\sigma^2} + \frac{2 \sigma^2 \lambda(k) + \mu^2}{\sigma^4} \right) \right],
\label{airy}
\ee
while $\lambda(k)$ is given by the largest root $\lambda$ of the following transcendental equation:
\begin{multline}
\left(-\frac{2k}{\sigma^2}\right)^{1/3} \Ai'\left[\left(-\frac{\sigma^2}{2k}\right)^{2/3}\frac{2 \sigma^2 \lambda + \mu^2}{\sigma^4} \right] \\
    \quad + \frac{\mu}{\sigma^2} \Ai \left[\left(-\frac{\sigma^2}{2k}\right)^{2/3}\frac{2 \sigma^2 \lambda + \mu^2}{\sigma^4} \right]=0.
\label{h2}
\end{multline}
As before, we can solve this equation numerically for given values of $\mu$ and $\sigma$ as well as various values of $k$ so as to build an interpolation of $\lambda(k)$, from which we obtain the rate function by computing the Legendre transform \eqref{eqlf1}. The resulting functions are shown in Figs.~\ref{fig:RBMDLD}. Unlike the ROUP, $\lambda(k)$ is now defined only for $k\leq 0$ because the potential
\be
V_k(x) = \frac{\mu^2}{2\sigma^2} - kx,
\ee 
which is related to the classic quantum triangular well, is confining only for $k < 0$, and so has bound states only for this range of parameters. This implies that the Legendre transform of $\lambda(k)$ gives $I(s)$ only for $s\in (0,s^*]$, as shown in Fig.~\ref{fig:RBMDLD}, where $s^*$ is again the typical value of $S_T$, found here from \eqref{eqgibbs1} and \eqref{eqcstar1} to be $s^*= \sigma^2/(2\mu)$.  

\begin{figure}[t]
\centering
\includegraphics{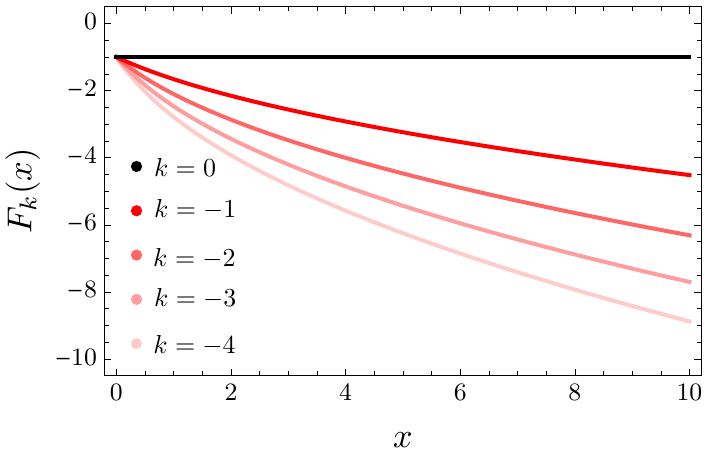}
\caption{(Color online) Driven force $F_k(x)$ for the RBMD with linear observable for various values of $k$. Parameters: $\sigma=1$ and $\mu=1$.} 
\label{fig:RBMDdrivenf}
\end{figure}

Above this value, $S_T$ does have fluctuations, but its large deviations are not covered by the spectral calculation, which is a sign generally that $P_T(s)$ scales weaker than exponentially in $T$. An example of such a scaling was discussed recently for the OUP \cite{nickelsen2018} using path integral techniques that predict a stretched exponential scaling in $T$, although the exact rate function cannot be found. The application of these techniques is beyond the scope of this paper, so we leave the study of the fluctuation region $S_T>s^*$ as an open problem \footnote{This region is not considered by Fatalov \cite{fatalov2017}.}. 

Note, incidentally, that for $\mu=0$, $S_T$ has no large deviations at the scale $T$ because Brownian motion is non-ergodic. The confining force produced by the negative drift along with the reflecting boundary makes that motion ergodic and creates fluctuations $S_T<s^*$ that are exponentially unlikely with $T$, though it is not strong enough, somehow, to constrain the fluctuations $S_T>s^*$ in the same exponential way. A similar effect is seen for the Brownian motion with reset \cite{hollander2018}.

To understand how the fluctuations of $S_T$ arise below its mean $s^*$, where $P_T(s)$ scales exponentially, let us analyze the driven process \footnote{Defining the driven process for $S_T>s^*$ is also an open problem, related again to the fact that the large deviations in that region are not given by the spectral elements of $\cL_k$ \cite{nickelsen2018}.}. The explicit expression of $F_k(x)$ for the RBMD is too long to show here, so we provide only the formula
\be
F_k(x)= \sigma^2\frac{\psi_k'(x)}{\psi_k(x)},
\ee 
which follows from \eqref{eqddf1} and \eqref{psirelate}, and which leads with \eqref{airy} to a ratio of the derivative of the Airy function and the Airy function. The result, plotted in Fig.~\ref{fig:RBMDdrivenf} for various values of $k$, is similar to the driven force found for the ROUP when $k<0$. Its shape shows that small fluctuations of $S_T$ are created, as expected, by squeezing the process close to the reflecting boundary by two forces: the negative constant drift $-\mu$ of the original RBMD and an added nonlinear force, which can be approximated near $x=0$ as a linear force with varying ``friction'' coefficient $\gamma_k =-F'_k(0^+)$. The action of these two forces creates a stationary density $\rho_k^*$ (not shown) which is approximately a shifted Gaussian density truncated at the boundary and increasingly concentrated there as $k \rightarrow -\infty$ or, equivalently, $S_T\ra 0$ \cite{buisson2020}. 

These results suggest that we approximate $F_k(x)$ for $k\leq 0$ to first order in $x$ as
\be
\tF_\theta(x) = -\theta x-\mu,
\ee
where $\theta \geq 0$. This has the same form as the ansatz \eqref{eqansatzforce2} used for the ROUP, with obvious replacements for the parameters, so we can find $\trho_\theta^*$ and $s_\theta$ from the results obtained before. The integral \eqref{eqcost1} of $K_\theta$, however, is different because of the different $F$ for the RBMD and leads now to
\be
\tI(s_\theta)=\left(\frac{\theta}{4}+\frac{\mu^2}{2\sigma^2}\right)-\frac{\mu}{2\sigma}\sqrt{\frac{\theta}{\pi}}\frac{e^{-\mu^2/(\theta \sigma^2)}}{1-\erf[\mu/(\sqrt{\theta}\sigma)]}.
\ee
It can be checked that the limit $\theta\ra 0$ of this approximation gives $\tI(s_0)=0$ at $s_0=s^*$, as expected, whereas $\theta\ra \infty$ gives a scaling near $s=0$ similar to the ROUP, namely, $\tI(s)\sim\sigma^2/(8\pi s^2)$.

\section{Conclusion}

We have shown how reflecting boundaries enter in the calculation of large deviation functions describing the likelihood of fluctuations of integrated quantities defined for Langevin-type processes. We have illustrated with basic examples the influence of such boundaries, particularly at the level of the driven process, which provides a mechanism for understanding how large deviations arise in the long-time limit. Our results pave the way for studying more general diffusions that evolve in confined domains of $\reals^2$ or $\reals^3$ with reflecting boundaries, as well as other observables, including particle currents, work-like quantities, and the entropy production, which are defined in terms of the increments of $X_t$ in addition to $X_t$ \cite{seifert2012}. The study of such diffusions should bring many new interesting results, as they may have non-zero stationary currents \cite{hoppenau2016} circulating parallel to a reflecting surface. Observables involving increments of $X_t$ are also expected to change the boundary term in the duality between $\cL_k$ and $\cL_k^\dag$ in a non-trivial way, giving rise to more complicated boundary conditions for $r_k$ and $l_k$.

In principle, our results can also be applied, as mentioned, to SDEs with multiplicative noise, that is, SDEs in which the noise amplitude $\sigma$ depends on the state $X_t$. These often arise in diffusion limits of population models \footnote{Note that jump processes, often used for modelling population dynamics, have no boundaries -- they have states connected by transition rates that can be set in an arbitrary way.}, as well as in finance, and should be treated in the same way as described here \footnote{SDEs with multiplicative noise can also be treated in one dimension by transforming them to SDEs with additive noise using the Lamperti transformation \cite{pavliotis2014}.}, assuming that their boundaries are reachable and that they are ergodic. The geometric Brownian motion, for example, is such that $X_t\geq 0$ but the boundary $x=0$ cannot be reached in finite time \cite{borodin2015}, so it does not make sense to define reflections there. This process is also not ergodic, so we do not expect a priori large deviations to exist.

Similar considerations apply to other boundary types: they should be treated in the same way as reflective boundaries by considering the boundary term arising in the duality of $\cL_k$ and $\cL_k^\dag$. However, as for multiplicative SDEs, we must ensure that the process considered with its boundary behavior is ergodic, for otherwise the distribution of observables is not expected to have a large deviation form. This prevents us, in general, from considering absorbing boundary conditions, which lead (without re-entry) to singular distributions concentrated on boundaries and for which, therefore, observables do not fluctuate in the infinite-time limit.

To conclude, we remark that our results could be obtained in a different way by using the contraction principle, which establishes a link between the large deviations of empirical densities and sample means, and which effectively replaces the spectral problem studied here by a minimization problem \cite{chetrite2015}. The boundary conditions that must be imposed on the latter problem are discussed for reflected diffusions by Pinsky \cite{pinsky1985,pinsky1985b,pinsky1985c,pinsky1985d} and can be shown to be equivalent to the zero-current conditions imposed on $\rho^*_k$, which represents in the contraction principle the optimal stationary density leading to a given fluctuation of $S_T$. This follows by generalizing a previous equivalence established for unbounded diffusions \cite{chetrite2015}.

In principle, one could also approach the large deviation problem by expressing the expectation $E[e^{kTS_T}]$ of the SCGF as a path integral restricted on an interval. Such integrals have been studied in quantum mechanics in the context of the free quantum particle evolving on the half-line \cite{clark1980,carreau1990,farhi1990}, but they are not expected to be solvable, except for simple systems. In any case, the main property of $E[e^{kTS_T}]$ that underlies dynamical large deviations is its exponential behavior in $T$, which, as we know from the Feynman--Kac equation, is determined by the dominant eigenvalue of the tilted generator $\cL_k$. Path integral techniques only confirm this result and have proved to be useful in large deviation theory mainly when considering the low-noise limit.

\begin{acknowledgments}
We thank Emil Mallmin for useful comments. J.d.B.\ is also grateful to the Joubert family for an MSc Scholarship administered by Stellenbosch University.
\end{acknowledgments}

\bibliography{masterbib}

\end{document}